# MultiAgent Deep Reinforcement Learning-Driven Mitigation of Adverse Effects of Cyber-Attacks on Electric Vehicle Charging Station

Manoj Basnet, *Student Member, IEEE* and Mohd. Hasan Ali, *Senior Member, IEEE*

*Abstract*— An electric vehicle charging station (EVCS) infrastructure is the backbone of transportation electrification. However, the EVCS has myriads of exploitable vulnerabilities in software, hardware, supply chain, and incumbent legacy technologies such as network, communication, and control. These standalone or networked EVCS open up large attack surfaces for the local or state-funded adversaries. The state-of-the-art approaches are not agile and intelligent enough to defend against and mitigate advanced persistent threats (APT). We propose the data-driven model-free distributed intelligence based on multiagent Deep Reinforcement Learning (MADRL)-- Twin Delayed Deep Deterministic Policy Gradient (TD3) -- that efficiently learns the control policy to mitigate the cyberattacks on the controllers of EVCS. Also, we have proposed two additional mitigation methods: the manual/Bruteforce mitigation and the controller clone-based mitigation. The attack model considers the APT designed to malfunction the duty cycles of the EVCS controllers with Type-I low-frequency attack and Type-II constant attack. The proposed model restores the EVCS operation under threat incidence in any/all controllers by correcting the control signals generated by the legacy controllers. Also, the TD3 algorithm provides higher granularity by learning nonlinear control policies as compared to the other two mitigation methods.

## I. Introduction

ACCORDING to the second-quarterly (Q2) data of 2021 from the Alternative Fuels Data Center, the United States hosts 128,474 public and private electric vehicle charging station (EVCS) ports in 50,054 different station locations [1]. In 2021 alone, charging stations increased by more than 55% in the United States. This upsurge is anticipated to grow further along with the announcement of the Bipartisan Infrastructure law to build out the nationwide electric vehicle network in April 2021 [2]. In February of 2022, USDOT and USDOE announced $5 billion over five years for the new National Electric Vehicle Under the Bipartisan Infrastructure law, the infrastructure (NEVI) program to create a network of EV charging stations and designated alternative fuel corridors on the interstate highway [3].

In contrast with the broad interest and investment in transportation electrification and EVCS deployment, the cyber-physical security hygiene of EVCS standalone/network is often slow-paced, poorly defined, and understudied [4]–[7]. The internet-facing elements of EVCS are primarily designed for communications and controls with other internet of things (IoTs) and stakeholders such as EV, EV operators, grid, Supervisory Control and Data Acquisition (SCADA), EVCS owners, and push the air-gapped critical physical infrastructures to the internet [8]. It could potentially open up large attack vectors for the interconnected systems of the EVCS.

Above all, there is an imminent need to develop distributed intelligence to proactively and independently defend the critical process controllers under the threat incidence. It motivates us to design, implement and test local, multiagent RL-based cyber defense that could detect and mitigate controller targeted advanced persistent threats (APT) in the EVCS charging process. De et al. designed a control-oriented model-based static detector (deviation in battery cell voltage) and a dynamic detector (using system dynamics) algorithms to detect denial of charging attacks and overcharging attacks on PEV battery packs [9]. The threshold-based static and filter-based dynamic detection techniques have the least flexibility toward APT and evolving zero-day attacks.

Girdhar et al. used the Spoofing, Tampering, Repudiation, Information Disclosure, Denial of service, Elevation of privilege (STRIDE) method for threat modeling and a weighted attack defense tree for vulnerability assessment in the ultra-fast charging ( XFC) station [10]. They proposed the Hidden Markov Model (HMM)-based detection and prediction system for a multi-step attack scenario. The proposed defense strategy optimizes the objective function to minimize the defense cost added by the cost of reducing the vulnerability index. As a means of defense/mitigation, the authors recommended isolating and taking the compromised EVCS off the interconnections and intercommunication. The traditional isolation-based protection approach miserably fails in the smart grid due to the availability constraints of electricity and few reserved physical backups. On this note, Mousavian et al. implemented the mixed-integer linear programming (MILP) that jointly optimizes security risk and equipment availability in grid-connected EVCS systems [11]. Still, their model aims to isolate a subset of compromised and likely compromised EVCS, ensuring the minimal attack propagation risk with a satisfactory level of equipment available for supply-demand. Acharya et al. derived the optimal cyber insurance premium for public EVCS to deal with the financial loss incurred by the cyberattacks [12].

In our previous work [7], we designed and engineered a Deep learning-powered (DNN, LSTM) network intrusion detection system that could detect DDoS attacks for the EVCS network based on network fingerprint with nearly 99% accuracy. Similarly, in [13], we developed a stacked LSTM-based host intrusion detection system solely based on a local electrical fingerprint that could detect stealthy 5G borne DDoS and FDI attacks targeting the legacy controllers of EVCS with nearly 100% accuracy. Furthermore, we proposed, tested, and evaluated several Deep Learning-based ransomware detection engines with the ability to share ransomware detection information in a cloud-based or distributed ransomware detection framework for EVCS [14].

Based on the above discussions, there are several significant findings: Firstly, state-of-the-art algorithms have progressed well for attack detection and prediction in EVCS, aided by cutting-edge computational intelligence at in-network and standalone levels. Secondly, CPS defense (capability of resisting attack) has been ill-defined and often confused with the mitigation (reducing the severity). As a result, CPS research is jumping towards mitigation that optimizes the cost function to protect remaining assets from further invasions by implementing predefined strategies, such as isolation of compromised assets, optimal insurance premium design, and mobilizing reserve resources. The obvious questions are, are we even trying to defend against any attacks in our critical CPS? Are we correcting the intruder/intruded signals? Thirdly, the current research lacks the convergence of IT security, OT security, and physical infrastructure security, with the slightest attention to OT and physical infrastructure security. The most devastating attacks in history have exploited the vulnerabilities in legacy controllers in OT or physical environments, as evident in Trident and Stuxnet attacks. Above all, the current state-of-the-art lacks a proactive vision for developing embedded intelligence that could defend/correct the attacks on the physical assets, mainly EVCS controllers.

To fill the gap in cyber-physical defense research at EVCS, we propose a novel, independent multiagent RL-based algorithm that oversees the critical functionality of all the controllers in the system and corrects and defend them under the detection of threat incidence as well as an anomaly. The proposed software agents operate solely based on the local data at EVCS to be purely air-gapped. Without shutting down the process, they can take over all the infected and frozen controllers under the worst cyberattack such as ransomware and/or APT. In addition, we have proposed two additional mitigation methods, the first with manual/bruteforce mitigation and the second with the concept of controller clone. We have compared and contrasted these mitigation measures with the proposed DRL-based mitigation. We have designed the PV-powered, off-the-grid standalone EVCS with a battery energy storage (BES) and an EV with the corresponding control circuitry of MPPT controller, PI controller-based BES controller, and EV controller. We carefully engineered and launched the APT attack (Type-I: low-frequency attack, Type-II: constant magnitude attack) on the duty cycle of the controllers. Since the scope of the paper is defense/mitigation, we have used a threshold-based detection engine for simplicity.

We summarize the contribution of this paper as follows:
- We proposed a novel data-driven multiagent TD3-based mitigation that could correct or take over the legacy controllers under the APT detection.
- Each agent is independent and can correct attacks on the corresponding legacy controller of EVCS. We envisioned the concept of embedded and distributed intelligence for critical legacy controllers.
- The proposed multiagents can learn and relearn the control policy online resulting from the change in dynamics of the EVCS environment or configurations. This is not the case in traditional controllers.
- The detection engine deploys the agent based on the infection of the legacy controller. The sophistication and accuracy of the detection engine can be upgraded by using AI.
- The proposed agents successfully restore the regular operation under the APT attacks and system anomaly on the legacy EVCS controllers.
- Finally, we compared the strengths and weaknesses of the proposed mitigation measures.

## II. SYSTEM MODEL AND MATHEMATICAL FORMULATION

Our previous work [14] designed the PV-powered off-the-grid standalone EVCS Prototype comprised of PV, BES, and EV with an associated control strategy. Fig. 1 depicts the SCADA system communicating with three isolated field controllers, namely PV, BES, and EV. The details about the EVCS architecture, control circuitry, system formulation, and component modeling are available in [13]. These field controllers are responsible for the reliable and safe operation of EVCS and hold the exploitable technical vulnerabilities. Using social engineering and/or reverse engineering, the adversary can poison the control signals reaching the physical controllers at EVCS either at the network level of the SCADA system or at the physical infrastructure layer. On that note, the threat actors with domain expertise can launch vicious APT attacks on these legacy controllers. To deal with these APT, Reinforcement Learning can be the reasonable control paradigm.

### A. Reinforcement Learning

Reinforcement learning is a goal-directed, direct, adaptive control that maps observation into actions to maximize the expected scalar reward founded on the notion of trial-and-error search and delayed reward [15]. Fig.1. depicts the working mechanisms of an individual DRL-based controller agent in EVCS and is valid for all controller agents: PV, BES, and EV agents. The detailed functionality and deployment of these agents with states, rewards, and actions information will be discussed in section IV.

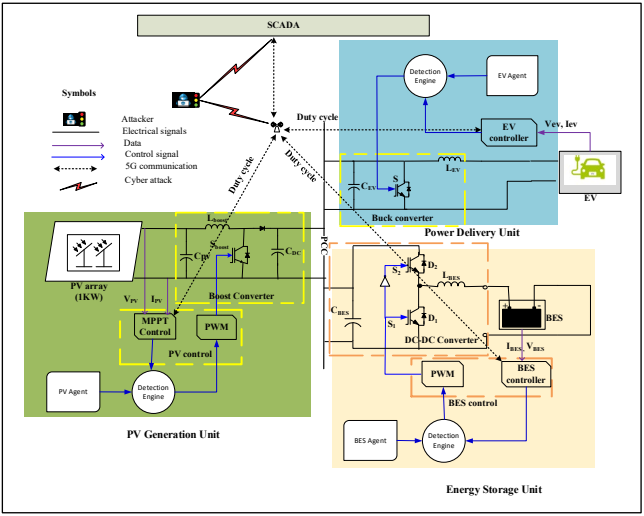

Fig. 1. Proposed Detection and Defense based on DRL based TD3.

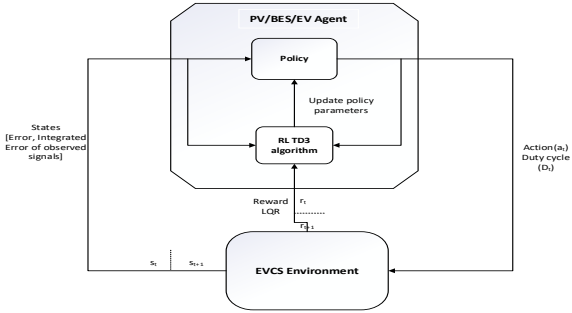

Fig. 2. The agent environment interaction of RL in a Markov Decision Process.

As per Fig. 2., at each discrete time step t, given the observation states $s_t \in S$ from EVCS, the agent (PV, BES, EV agent) selects actions $a_t \in A$ with respect to policy $\pi: S \to A$ receiving a reward $r_t$ and new state of the environment $s_{t+1}$. Where $S$ is the set of {error, integrated error} of the critical electrical parameter of EVCS having an immediate impact on the control action $A$, i.e., duty cycle. The discounted sum of rewards $R$ is given in (1), where the discount factor $\gamma \in [0,1]$ represents the priority of short-term rewards.

$$R \triangleq \sum_{i=t}^{T} \gamma^{i-t} r(s_i, a_i) \quad (1)$$

Markov Property: Given the present, it states that the future is independent of the past. A state $s_t \in S$ is Markov if and only if

$$P[s_{t+1}|s_t] \triangleq P[s_{t+1}|s_1, \ldots, s_t] \quad (2)$$

### B. Markov Decision Process (MDP)

MDP formally characterizes an environment in an RL where the current state describes the process completely. An MDP is a tuple of $M = \langle S, A, P, R, \gamma \rangle$ Where $S$ is a finite set of states, $A$ is a finite set of actions, and $P$ is a state transition probability referring to the likelihood of going to the next state $s'$ from the current state $s$ after taking action $a$ defined in (3).

$$P_{ss'} = P[s_{t+1} = s'|s_t = s, a_t = a] \quad (3)$$

The goal of reinforcement learning is to find the optimal policy $\pi_\phi$ with parameter $\phi$ that maximizes the expected return as in (4). The parameterized policy $\pi_\phi$ for continuous control can be updated by taking the gradient of the expected reward $\nabla_\phi J(\phi)$ as in (5). The actor or policy can be updated using a deterministic policy gradient algorithm in Actor-Critic methods. The critic or the value function, $Q^\pi(s, a)$, is defined as the expected return while taking action $a$ in state $s$ and following the policy $\pi$ after that as in (6). One of the ways to learn the value function is to use a temporal difference learning in the Bellman equation that relates the value of the current state-action pair to the value of the next state-action pair as (7). As there are too many states and actions to store in memory in the large MDP, the value function can be approximated by using a differentiable function approximator $Q_\theta(s, a)$. It is parameterized by $\theta$ as in (8). The parameter $\theta$ can be updated by using Monte Carlo or Temporal difference learning. In this learning of DQN, $\theta$ is updated by using a secondary frozen target network $Q_{\theta'}(s, a)$ to maintain the fixed objective $y$ over multiple updates as in (9). The actions $a'$ are selected from the target actor-network $\pi_{\phi'}(s')$. The parameter $\theta'$ of the target network can be updated by exactly matching the weight of the current network or by some proportion $\tau$ at each timestep as in (10).

$$J(\phi) = \mathbb{E}_{s_i \sim p_\pi, a_i \sim \pi}[R_0] \quad (4)$$

$$\nabla_\phi J(\phi) = \mathbb{E}_{s \sim p_\pi}[\nabla_a Q^\pi(s,a)|_{a=\pi(s)} \nabla_\phi \pi_\phi(s)] \quad (5)$$

$$Q^\pi(s, a) = \mathbb{E}_{s_i \sim p_\pi, a_i \sim \pi}[R_t | s, a] \quad (6)$$

$$Q^\pi(s, a) = r + \gamma \mathbb{E}_{s',a'}[Q^\pi(s', a')], \ a' \sim \pi(s') \quad (7)$$

$$Q_\theta(s, a) \approx Q^\pi(s, a) \quad (8)$$

$$y = r + \gamma Q_{\theta'}(s', a'), \ a' \sim \pi_{\phi'}(s') \quad (9)$$

$$\theta' \leftarrow \tau\theta + (1-\tau)\theta' \quad (10)$$

### III. ATTACK MODELING AND DETECTION

#### A. Attack Modelling

The attacker's primary goal is to disrupt, damage, or freeze the critical controllers of EVCS. The attacker is assumed to poison/manipulate the critical parameters with sophisticated attacking tools and domain expertise. Most legacy controllers generate a critical control signal, i.e., the duty cycle that controls the switching of the high-frequency transistor switches. It is assumed that the attacker can control the number of controllers ($N_c \in \mathbb{R}$), the attack duration ($T_a \in \mathbb{R}$) and Types of the attack $S_a = \{(A_t, E_a)\}$ once it exploited the critical control signals $C = \{C_1, C_2, \ldots, C_n\}$ from controllers $N_1, N_2, \ldots, N_n$. The attacker chooses the set of exploited resources $\zeta$ from another set $\mathcal{M} = \{N_c, T_a, S_a, C\}$ in such a way to minimize the critical functionality $CF$ of the process as in (11). The attack Type $S_a$ can be a tuple of attack time $A_t =$

$\{sim, diff\}$ and engineered attack Types $E_a = \{\tau_1, \tau_2\}$ where *sim* and *diff* refer to the attack that can be launched simultaneously and at different times, respectively, with attack Types $\tau_1$ and $\tau_2$ defined in (12) and (13).

$$\underset{\zeta \subset \mathcal{M}}{\operatorname{argmin}} \, CF \qquad (11)$$

$$\tau_1 = T(\alpha) \qquad (12)$$

$$\tau_2 = c \qquad (13)$$

Where T is some random function parameterized by parameters $\alpha$ and c is some scalar constant. The function T is envisioned to generate the statistical randomness in the attack. With critical control signals of the controllers $\mathcal{C} \in [low\_limit, upper\_limit]$, it is wise for a stealthy attacker to design a similar kind of pseudorandom attack that intersects with the range of $\mathcal{C}$. Pseudorandom number *PRN(low_limit, up_limit, rep)* fluctuates between lower and upper bound and repeats *rep* times serve the purpose. Similarly, c is the average of the upper and lower limit of the $\mathcal{C}$. After finding the sets of optimal $\zeta$, Finally, the attacker algebraically combines the attack signal $E_a$ with the critical parameter set $\mathcal{C}$ as per (14).

$$attack\_signal = \mathcal{C} \pm E_a \text{ subjected to } \zeta \qquad (14)$$

TABLE I
PARAMETERS FOR ATTACK MODELING

| Parameters | Value |
|---|---|
| $N_c$ | 3 |
| $T_a$ | 2 seconds |
| $S_a$ | $\{sim, diff\} \times \{\tau_1, \tau_2\}$ |
| $\mathcal{C}$ | $\{\mathfrak{D}_{PV}, \mathfrak{D}_{BES}, \mathfrak{D}_{EV}\}$ |
| CF | Observed normalcy or stability of the process variables such as power, bus voltage, BES, and EV voltages and currents |
| $\tau_1$ | PRN(0, 1,10) |
| $\tau_2$ | 0.5 |
| Type I attack = $\mathcal{C} \pm E_a$ s.t. $\zeta(.\tau_1)$ | $\{\mathfrak{D}_{PV} + \tau_1, \mathfrak{D}_{BES} - \tau_1, \mathfrak{D}_{EV} - \tau_1\}$ |
| Type II attack = $\mathcal{C} \pm E_a$ s.t. $\zeta(.\tau_2)$ | $\{\mathfrak{D}_{PV} - \tau_2, \mathfrak{D}_{BES} - \tau_2, \mathfrak{D}_{EV} - \tau_2\}$ |

We pragmatically chose the $\zeta$ and $CF$ of the attack for this case as in table I after the repeated experimentation. Both Type I and Type II attacks are carefully engineered APT attacks with domain expertise. The Type I attack imposes the low-frequency attack on the duty cycles, while the Type II attack imposes the constant duty cycle attack.

*B. Detection Technique*

The threshold-based detection engine is a good choice for simplicity and speed as it makes the online decision after implementing rule-based logic. This detection engine is founded on point anomaly detection that continuously oversees whether the control signal and controlled signal are within the predefined threshold range. If these signals fall outside the predefined thresholds, the engine decides it as an anomaly or attack. The detection logic at each controller can be defined as follows.

If ($Upper\ Threshold < \mathfrak{D}_{()} < Lower\ Threshold$) && ($Upper\ Threshold < CF_{()} < Lower\ Threshold$)
  Continue with Legacy controllers, i.e., duty=$\mathfrak{D}_{()}$;
Else
  Correct the duty cycle with TD3 agent, i.e., duty =$\mathfrak{D}_{(corrected)}$;
end

The complete information of thresholds for respective controllers is presented in section IV, parts D, E, and F, along with the agent design.

IV. PROPOSED MITIGATION APPROACH

We proposed the multiple independent DRL-based data-driven TD3 agent trained and deployed in each critical controller that controls a dynamic system's critical functionality. Upon the threat incidence or operational anomaly, the rule-based or DL-based detection engine deploys the corrected control signal from agents. It takes over the legacy controllers until the threat has been eliminated. We have proposed an RL-based autonomous defense agent for each controller whose primary purpose is to generate the corrected control signal upon incidences of cyberattacks and system anomalies. These controllers are designed for the mere to extreme adversarial setups such as APT or malware that could freeze/control the legacy controllers.

*A. Twin Delayed Deep Deterministic Policy Gradient (TD3)*

Actor-critic RL learns both value function (as in value-based RL) and policy (as in policy-based RL) and is proven with better convergence properties, ability to learn stochastic policy, and efficacy in hyperdimensional or continuous action space. The function approximation error in actor-critic RL leads to overestimated value estimates and suboptimal policies [16]. TD3 is the off-policy actor-critic RL designed for continuous action space to minimize the impact of overestimation bias on both actor-critic networks by implementing three tasks. The first one is Clipped Double -Q Learning, where TD3 uses a minimum of two Q-values to form the target in (7). The second one is the delayed policy updates of the target network. And the third one is the Target policy smoothing, where TD3 adds noise to the target action so that the target policy could not exploit Q function error by smoothing out Q along the gradient of action.

*B. Training Algorithm for Twin Delayed DDPG(TD3)*

Each proposed standalone control agent for EVCS follows the strict training protocol as follows.

Initialize critic networks $Q = [Q_{\theta_1}, Q_{\theta_2}]$ and actor-network $\pi_\phi$ with random parameters $\theta_1, \theta_2$ and $\phi$
  Initialize target networks $\theta'_1 \leftarrow \theta_1, \theta'_2 \leftarrow \theta_2, \phi' \leftarrow \phi$
  Initialize replay buffer $\mathcal{B}$
  for *t*=1 to *T* do
    Select action with exploration noise $a \sim \pi_\phi(s) + \epsilon$ where $\epsilon \sim \mathcal{N}(0, \sigma)$
    Store transition tuple $\langle s, a, r, s', d \rangle$ into $\mathcal{B}$ where *d* is the signal to indicate $s'$ is the terminal state
    If $s'$ is the terminal state, reset environment state
    Else randomly sample mini-batch of *N* transitions

⟨s, a, r, s′, d⟩ from $\mathcal{B}$

Compute the target actions and compute targets:

$a'(s')$
$= clip(\mu_{\phi'}(s') + clip(\epsilon, -c, c), a_{low}, a_{high}), \epsilon \sim \mathcal{N}(0,1)$

$y(r, s', d) = r + \gamma(1-d) \min_{i=1,2} Q_{\theta'_i}(s', a'(s'))$

Update critics Q-function by using one step of gradient descent:

$\theta_i \leftarrow \operatorname{argmin}_{\theta_i} \nabla_{\theta_i} \frac{1}{|\mathcal{B}|} \sum_{\langle s,a,r,s',d \rangle \in \mathcal{B}} (Q_{\theta_i}(s,a) - y(r,s',d))^2$

**If** $t$ mod *policy_delay*, **then**

Update $\phi$ by the deterministic policy gradient:

$\nabla_\phi J(\phi) = \frac{1}{|\mathcal{B}|} \sum \nabla_a Q_{\theta_i}(s,a)|_{a=\mu_\phi(s)} \nabla_\phi \pi_\phi(s)$

Update target networks:
$\theta'_i \leftarrow \tau\theta_i + (1-\tau)\theta'_i$
$\phi' \leftarrow \tau\phi + (1-\tau)\phi'$
**End if**
**End for**

### C. Graphical representation of TD3 algorithm

TD3 uses twin critic networks (critic 1 and critic 2) inspired from DRL with clipped Double Q-Learning, where it takes the smallest Q-value of two critic networks to remove the overestimation bias in $Q_{\theta_i}(s,a)$. The concept of target networks is introduced to stabilize the agent training. The target network provides a stable objective and greater coverage of the training data as DNN requires multiple gradient updates to converge [16]. Without the fixed target, the accumulated residual errors after each update produce divergent values when paired with a policy maximizing the value estimate. Therefore, TD3 uses delayed update of actor-network (policy update) compared to critic network (value update), resulting in more stable training.

The replay buffer stores the history of agent experience and randomly fetches the data in mini-batches to update actor and critic networks. There are six neural networks in TD3: two critic networks, two target networks for two critics, an actor-network, and a corresponding target network for the actor. Fig. 3 summarizes the graphical abstract of a TD3 agent.

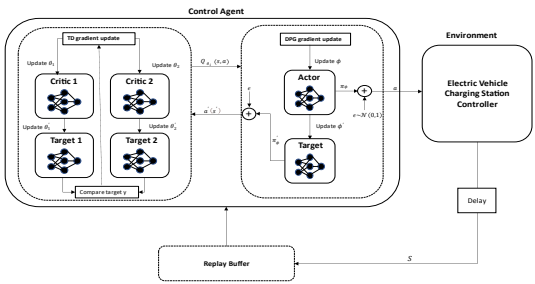

Fig. 3. Graphical representation of a TD3 agent.

### D. PV agent

The design goal of the PV agent is to take over the infected MPPT controller and implement the optimized control policy to generate the duty cycle $\mathfrak{D}_{PV}(t)$ needed by the boost converter to have the least impact on the system. A PV agent continuously monitors the error $e_P(t)$ and integrated errors $e_{int\_P}(t)$ between the PV output power $P_{PV}(t)$ and reference power $P_{ref}(t)$ as in (15) and (16). The objective of the PV agent is to find the optimal policy for the duty cycle that correctly transforms observation space to action space by maximizing the cumulative scalar reward.

The output or action of the PV agent is the duty cycle with a linear quadratic regulator (LQR) as the instantaneous reward or cost function $r_{PV}(t)$ as in (17). The $\alpha = 0.01$ and $\beta = 1$ on $r(t)$ represent the negative penalty terms imposed on error and action, respectively. $T_S$ is the sampling time and is the same for each agent.

$$e_P(t) = P_{ref}(t) - P_{PV}(t) \tag{15}$$

$$e_{int\_P}(t) = \sum_{T_S} e_P(t)) \tag{16}$$

$$r_{PV}(t) = \alpha e_P(t)^2 + \beta \mathfrak{D}_{PV}(t)^2 \tag{17}$$

The rule/threshold-based detection engine derived pragmatically for PV agent will determine the attack event if observed power falls beyond the range (1020,1045) AND duty of MPPT falls beyond the range (0.200,0.201).

### E. BES agent

The design goal of the BES agent is to generate the corrected duty cycle $\mathfrak{D}_{BES}(t)$ for the buck-boost converter under the threat incidence. Similar to the PV agent, the BES agent observes the states of the error $e_V(t)$ and integrated errors $e_{int\_V}(t)$ between the desired reference bus voltage $V_{bus\_ref}(t)$ and the bus voltage $V_{bus}(t)$ as in (18) and (19). The optimal control policy that maps the observation space to action space is found by minimizing the expected value of the cost function $r_{BES}(t)$, which is the linear quadratic regulator function. The $\alpha = 0.01$ and $\beta = 1$ on $r_{BES}(t)$ represent the negative penalty terms imposed on error and action, respectively as in (20).

$$e_V(t) = V_{bus\_ref}(t) - V_{bus}(t) \tag{18}$$

$$e_{int\_V} = \sum_{T_S} e_V(t)) \tag{19}$$

$$r_{BES}(t) = \alpha e_V(t)^2 + \beta \mathfrak{D}_{BES}(t)^2 \tag{20}$$

The rule/threshold-based detection engine derived pragmatically for the BES agent will determine the attack event if observed power falls beyond the range (1020,1045) and the PI controller's duty falls beyond the range (0.7,0.71).

### F. EV agent

The design goal of the EV agent is to generate the corrected duty cycle $\mathfrak{D}_{EV}(t)$ for a buck converter if the legacy EV charger got infected. Similar to the previous agent, the EV agent observes the states of the error $e_{VEV}(t)$ and integrated errors $e_{int\_VEV}(t)$ between the desired reference battery voltage $V_{batt\_ref}(t)$ and the bus voltage $V_{batt}(t)$ as in (21) and (22). The optimal control policy that maps the observation space to action space is found by minimizing the expected value of the cost function $r_{EV}(t)$, which is the linear quadratic regulator function. The $\alpha = 0.01$ and $\beta = 1$ on $r_{EV}(t)$ represent the negative penalty terms imposed on error and action, respectively as in (23).

$$e_{VEV}(t) = V_{batt\_ref}(t) - V_{batt}(t) \tag{21}$$

$$e_{int\_VEV} = \sum_{T_s} e_{VEV}(t)) \tag{22}$$

$$r_{EV}(t) = \alpha e_{VEV}(t)^2 + \beta \mathfrak{D}_{EV}(t)^2 \tag{23}$$

The rule/threshold-based detection engine derived pragmatically for the EV agent will determine the attack event if observed power falls beyond the range (1020,1045) and the PI controller's duty falls beyond the operating range (0.54,0.55).

Table II summarizes the observations, reward, and action information of multiple TD3 agents.

TABLE II
SUMMARY OF MULTIPLE INDEPENDENT AGENTS

| Agents | Observations ($\mathcal{S}$) | Reward ($\mathcal{R}$) | Action ($\mathcal{A}$) |
|---|---|---|---|
| PV Agent | $\{e_P, e_{int\_P}\}$ | $\{r_{PV}\}$ | $\{\mathfrak{D}_{PV}\}$ |
| BES Agent | $\{e_V, e_{int\_V}\}$ | $\{r_{BES}\}$ | $\{\mathfrak{D}_{BES}\}$ |
| EV Agent | $\{e_{VEV}, e_{int\_VEV}\}$ | $\{r_{EV}\}$ | $\{\mathfrak{D}_{EV}\}$ |

### G. Additional Cyberattack Mitigation Methods

Along with the TD3-based mitigation, we devised two additional cyberattack mitigation methods for EVCS: a heuristic-based Manual/Brute force mitigation and a controller clone-based mitigation. These two methods are explained below in parts G.1 and G.2.

#### G.1) Manual/Brute force mitigation

The brute force mitigation is the manual override of the control signal under cyber-attack detection. This approach might be good for the stable and convergent linear time-invariant system. In this approach, the critical control signals are learned from the analysis of repeated experimentations with human domain experts. These learned signals are released as a corrective measure to the physical controllers under the threat incidence. The working logic of brute force mitigation is given below.

If a duty cycle attack is detected in a controller, release heuristic-based predefined duty cycle

Else continue with legacy controllers.

After careful examination of the process, the desired control signals, the duty cycle for PV, BES, and EV controller, are 0.2, 0.7, and 0.55, respectively, for the stable operation of EVCS. This method can restore the EVCS operation under a constrained environment, with the following limitations.

- This method is purely static and not intelligent (it does not have learning capability); therefore, it is not adaptive.
- Small changes in operational conditions or minor flaws can fail the model, i.e., very high failure susceptibility.
- This model is prone to failure under the APT or ransomware attack that can freeze the controllers where manual overriding is not an option anymore.

#### G.2) Controller clone-based mitigation

We develop the concept of controller twins to deal with the freezing of controllers due to the worst-case cyberattacks such as APT and Ransomware. Unlike brute force mitigation, this model has an exact clone of the controllers, meaning the same operational technologies and configurations in case one fails; the clone can take over. The control logic of this method is given below.

If a duty cycle attack is detected in a controller, replace the infected controller with its physical clone

Else continue with legacy controllers.

For the EVCS, exact physical copies of controllers with the same configurations and operating principles are deployed under the attack detection on the operating controllers. This method outperforms the manual brute force-based control. However, this method has the following limitations.

- The controller twins share the same vulnerabilities as the operating ones; hence, they can be easily exploitable.
- These are adaptive; however, they are not intelligent enough to have learning capability.
- It can't learn the nonlinear control policy.
- Changes in operational setpoints and configurations update need retuning of the controllers.
- Very high economic overhead is needed to set up the entire clone of the operating controllers.

## V. EXPERIMENTAL SETUPS FOR TD3-BASED METHOD

The TD3-based multiagents are built with specific neural architectures for critics and actor neural networks with similar architecture for the target neural network. Then, the layerwise actors and critics neural networks with their targets are properly engineered and parameterized with desired activation functions and appropriate initial weights and biases. Finally, the hyperparameters are carefully selected to train the agents in an optimized way.

### A. Configurations of TD3 Critic networks

A TD3 critic estimates the optimal Q-value based on the observations and actions received by the parameterized DNN. Fig. 4 depicts the structure of a single critic network we have created. Before concatenating those features, the state and action information goes through some local neural network transformations. After concatenation, it goes through another set of neural networks to produce the Q-value function. The network that takes state info has three fully connected hidden layers with respective hidden units of 64, 64, 32, and the relu activation layer between them. Also, the action info is passed through the fully connected neural network with 64 hidden units. After the concatenation, the transformed state and action info pass-through two fully connected hidden layers with 64 and 32 hidden units, respectively, with relu layer in between to produce the Q-value. We then create the critic representation using specified neural networks and options.

## B. Configurations of TD3 Actor networks

The actor-network in the TD3 agent decides which action to take based on the observations, i.e., policy mapping. We have created a DNN with three fully connected hidden layers with respective hidden units of 64, 32, and units equal to the number of actions, i.e., 1 in our case with relu layers in between. In addition, a sigmoid layer is added since the output of the action ranges from 0 to 1 for duty cycle in our case. Finally, the scaling layer scales the output from the sigmoid layer with a scale of 1 and a bias of 0.5. The scale is set to the range of action signal, and bias is set to with half a range. We then create the actor representation using specified neural networks and options as in Fig. 5. Table III presents the options of actor-network, critic network as well as training of agent. Table IV presents the hyperparameters setting to administer the training.

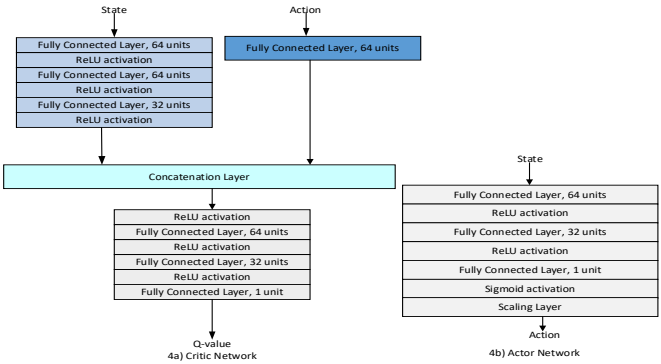

Fig. 4. Structure of proposed a) Critic-Network b) Actor-Network.

TABLE III
ACTOR-CRITIC NETWORK PARAMETERS

|        | Optimizer | Learning Rate | Gradient Threshold | L2 Regularization Factor |
|--------|-----------|---------------|---------------------|--------------------------|
| Critic | Adam      | 0.001         | 1                   | 0.0002                   |
| Actor  | Adam      | 0.001         | 1                   | 0.00001                  |

TABLE IV
TRAINING PARAMETERS SETTING

| Discount factor | 0.99 |
|---|---|
| Experience buffer Length | $10^6$ |
| Mini-batch size | 128 |
| Number of steps to look ahead | 10 |
| Target smooth factor | $5e^{-3}$ |
| Target update frequency | 2 |
| Exploration variance | 0.01 |
| Target Policy smooth variance | 0.2 |

## C. Training an agent

The agent trains by randomly selecting mini-batches of size 128 with a discount factor of 0.99 towards the long-term reward from the replay buffer or experience buffer with a maximum capacity of 1e6. The target critics and actor are time-delayed clones of the critics and actor network with a smoothing factor of 0.005 that update every 2 agent steps during training. The agent uses a Gaussian action noise model with a specified noise variance and decay rate to explore the action space during training. The agent also smooths the target policy updates using the specified Gaussian noise model.

Each training consists of 200 episodes, with each episode consisting of nearly 170 steps. The agent training is terminated when the agent receives an average cumulative reward of more than 800 over 100 consecutive episodes.

## D. Computational performance

The proposed RL TD3 agents, namely: the PV agent, BES agent, and EV agent, train individually as the agents should learn to act independently. The motive behind designing the independent multiagent is that they should be able to work with legacy controllers ( in case only a few controllers got infected) and other trained RL TD3 agents (all legacy controllers got infected). We train the agents as configured in sections IV D), E), and F) independently.

Since we are about to engineer the independent agents, there won't be a collaborative or adversarial learning paradigm such as the concept of hierarchical (global and local) rewards. Instead, we retrain the EV agent with the trained RL TD3 PV and BES agents so that it can upgrade the learned policy. After that, we test various combinations of RL TD3 agents and legacy controllers to assess the control actions on the EVCS. Finally, the trained agents are deployed with the tendency to upgrade the policy while EVCS is running. All the computations and simulations are performed in Matlab 2021 b and Simulink 3.253 R2021b in Dell XPS 15 7590 machine with i7-9750H CPU @2.6 GHz and 16GB RAM. Each agent took approximately 6.68 hours of training for the 200 episodes. The training progress in terms of rewards is shown in Fig. 5 with stabilized reward within 103 episodes for PV agent, 105 episodes for BES agent, and 101 episodes for EV agent.

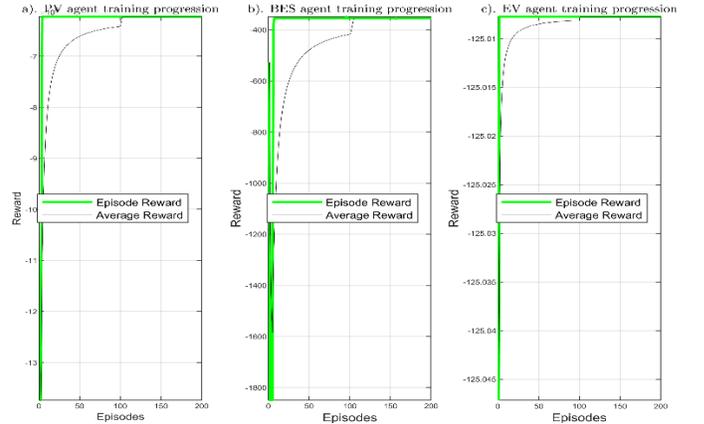

Fig. 5. Training performance of the Agents.

## VI. SIMULATION RESULTS AND DISCUSSION

### A. Type I and Type II Attacks

Fig. 6 summarizes the Type I and Type II attacks launched at different controllers at the same time and at different times. The Type I attack is the low-frequency attack at the duty cycles of the controllers, while Type II is the constant attack. The BES duty cycle is found to be more vulnerable to both kinds of attacks than the duty cycles of other controllers. The Type I attack has an irreversible impact on the BES controller as opposed to the Type II attack on the BES controller and both attacks on other controllers.

## B. Type I Attack on Different Times and Mitigation Analysis

The Type I attack has been launched in three different controllers PV controller at 5-7 seconds, BES controller from 9-11 seconds, and EV controller from 13-15 seconds, as shown in Fig. 7. Tables V, VI, VII, VIII, IX, and X present the corresponding statistics of important electrical parameters.

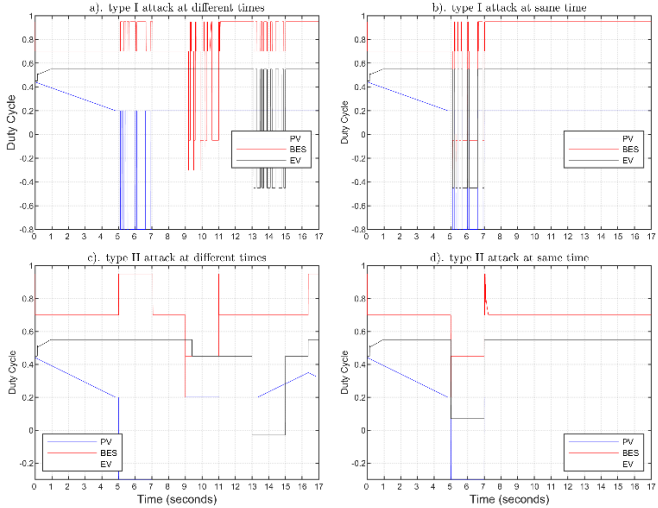

Fig.6. Impacts of the Type I and Type II attacks on duty cycles of different controllers.

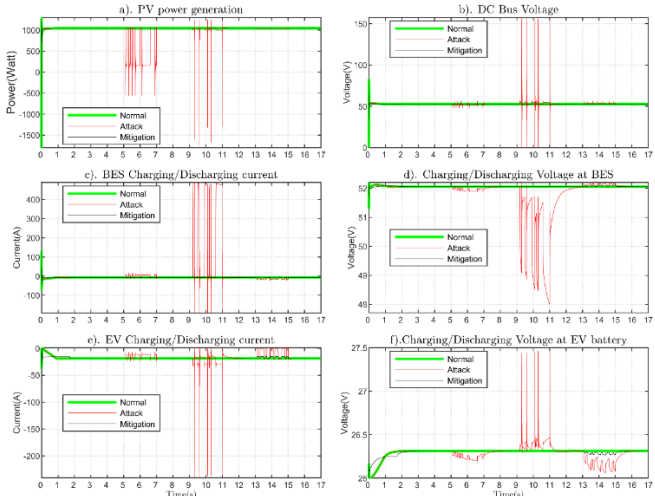

Fig. 7. Impacts of Type I attack launched at PV controller from 5-7 seconds, BES controller from 9-11 seconds, and EV controller from 13-15 seconds and the mitigation performance during the attack

The Type I attack has impacted all the critical electrical parameters. It forces the power to have approx. 2.99k times the normal range, 7.5k times the normal interquartile range (IQR), and median less than 18.4 Watt to the median at regular operation. The proposed mitigation restores the power with approximate errors of 0.002 watts in the median, 0.0001 watts in IQR, and -2.44 watts in range with the one at normal operations, as evident in Table V.

Similarly, the Type I attack has an inverse impact on bus voltage with a range elevation of approximately 158 V, IQR elevation of 1.63 V, and median reduction by 0.0052 V compared to the base operating conditions. The proposed mitigation can restore the bus voltage with approximate errors of 0 V in the median, 0.0001 V in IQR, and 0.5288 V in the range with the one at normal operations as per Table VI.

TABLE V
PV POWER STATISTICS IN WATT DURING NORMAL, ATTACK AND MITIGATION

|  | Range | IQR | median |
|---|---|---|---|
| Normal | [1043.59,1044.60] | [1043.593,1043.599] | 1043.5996 |
| Attack | [1768.23,1255.32] | [998.97,1043.71] | 1025.1726 |
| Mitigation | [1040.15,1043.60] | [1043.594,1043.5998] | 1043.5969 |

TABLE VI
DC BUS VOLTAGE STATISTICS IN VOLTS DURING NORMAL, ATTACK AND MITIGATION

|  | Range | IQR | median |
|---|---|---|---|
| Normal | [52.7593, 52.761] | [52.7596,52.7607] | 52.7605 |
| Attack | [-0.23051,157.502] | [52.2102,53.8464] | 52.7553 |
| Mitigation | [52.608,53.1379] | [52.7596,52.7608] | 52.7605 |

Also, as per table VII, the Type I attack has an inverse impact on BES current with a range elevation of approximately 683 A, IQR elevation of 14 A, and median increment by 0.1 A compared to the base operating conditions. The proposed mitigation can restore the BES current with approximate errors of 0.0001 A in the median, 0.0013 A in IQR, and 2.4159 A in the range with the one at normal operations.

TABLE VII
BES CURRENT STATISTICS IN AMPERE DURING NORMAL, ATTACK AND MITIGATION

|  | Range | IQR | median |
|---|---|---|---|
| Normal | [-6.947,-6.9435] | [-6.9435,-6.944] | -6.9452 |
| Attack | [-193.294,489.396] | [-8.816,- 6.358] | -6.8273 |
| Mitigation | [-9.143,-6.723] | [-6.9456, -6.944] | -6.945 |

Likewise, the Type I attack has an inverse impact on BES voltage with a range elevation of approximately 4.3434 V, IQR elevation of 0.5747 V, and median decrement by 0.0914 V compared to the base operating conditions. The proposed mitigation can restore the BES current with approximate errors of 0.000 V in the median, 0.0102 V in IQR, and 0.0251V in the range with the one at normal operations evident from table VIII.

TABLE VIII
BES VOLTAGE STATISTICS IN VOLTS DURING NORMAL, ATTACK AND MITIGATION

|  | Range | IQR | median |
|---|---|---|---|
| Normal | [52.0586,52.0588] | [52.0586,52.0588] | 52.0587 |
| Attack | [47.9982,52.3418] | [51.5319,52.1066] | 51.9673 |
| Mitigation | [52.0586,52.0839] | [52.0586,52.069] | 52.0587 |

Table IX shows that the Type I attack has an inverse impact on EV current with a range elevation of approximately 241.2919 A, IQR elevation of 8.0385 A, and median increment by 2.0851 A compared to the base operating conditions. The proposed mitigation can restore the EV current with approximate errors of 3e-4 V in the median, 0.0015 A in IQR, and 4.0534 A in the range with the one at normal operations.

Table X implies that the Type I attack has an inverse impact on EV voltage with the range elevation of approximately 1.4074 V, IQR elevation of 0.1438 V, and median decrement by 0.0602 V compared to the base operating conditions. The proposed mitigation can restore the BES current with approximate errors of 4.0000e-04 V in the median, 0.0193 V in IQR, and 0.0472 V in the range with the one at normal operations.

TABLE IX
EV CURRENT STATISTICS IN AMPERE DURING NORMAL, ATTACK AND MITIGATION

|  | Range | IQR | median |
|---|---|---|---|
| Normal | [-18.674,-18.668] | [-18.674,-18.669] | -18.6713 |
| Attack | [-241.298,5.26e-5] | [-19.531, -11.488] | -16.5862 |
| Mitigation | [-18.9473,-14.887] | [-18.674,-18.671] | -18.6716 |

TABLE X
EV VOLTAGE STATISTICS IN VOLTS DURING NORMAL, ATTACK AND MITIGATION

|  | Range | IQR | median |
|---|---|---|---|
| Normal | [26.312,26.313] | [26.312,26.313] | 26.3126 |
| Attack | [26.056,27.465] | [26.199,26.344] | 26.2524 |
| Mitigation | [26.265,26.313] | [26.293,26.313] | 26.3122 |

### C. Type I Attack simultaneously on all controllers and mitigation analysis

The Type I attack has been launched simultaneously in three different controllers at 5-7 seconds as shown in Fig. 8. The Type II attack that launches at different times has impacted all the critical electrical parameters.

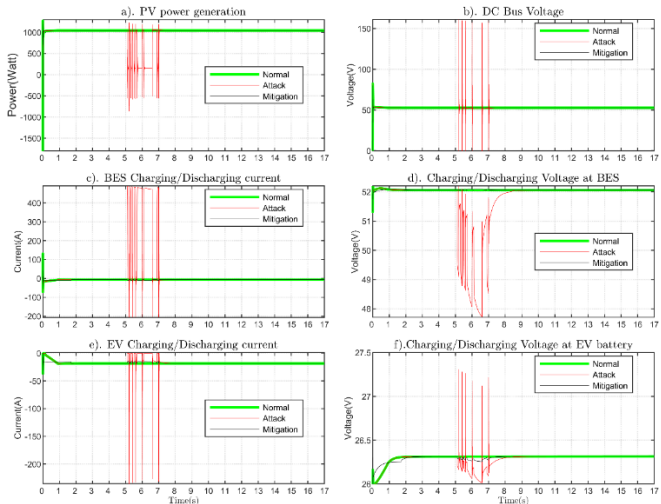

Fig. 8. Impacts of Type I attack launched at PV controller from 5-7 seconds, BES controller from 5-7 seconds, and EV controller from 5-7 seconds and the mitigation performance during the attack.

### D. Type II Attack on different times and mitigation analysis

The Type II attack has been launched in three different controllers PV controller at 5-7 seconds, BES controller from 9-11 seconds, and EV controller from 13-15 seconds, as shown in Fig. 9.

### E. Type II Attack simultaneously on all controllers and mitigation analysis

The Type II attack was launched simultaneously in three different controllers at 5-7 seconds as shown in Fig. 10.

### F. Performance Comparison of Various Proposed Methods

Fig. 11 depicts the control actions, i.e., duty cycles of three proposed methods under the Type-II attack at all controllers from 5-7 seconds. For the correction of the PV duty cycle, the manual mitigation provides the constant duty cycle of 0.2 no matter what happens to the process. In contrast, the controller clone settles at a duty cycle of 0.2 after nearly 5 seconds. However, the TD3 mitigation converges on an optimal duty cycle than the traditional MPPT controller and finely operates the EVCS system. Similarly, the TD3 algorithm has learned nonlinear optimal control policies for BES and EV control, instead of linear control of brute force and controller clone methods. Hence, the TD3 algorithm provides higher granularity by learning control policies that the proposed other two methods can't learn.

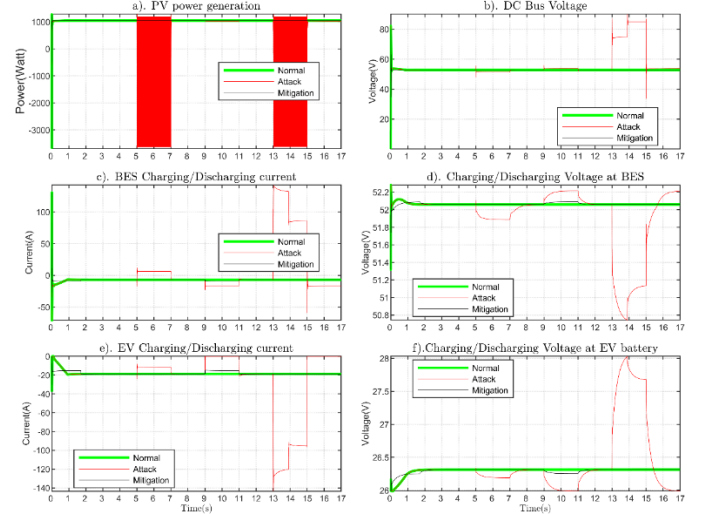

Fig. 9. Impacts of Type II attack launched at PV controller from 5-7 seconds, BES controller from 9-11 seconds, and EV controller from 13-15 seconds and the mitigation performance during the attack.

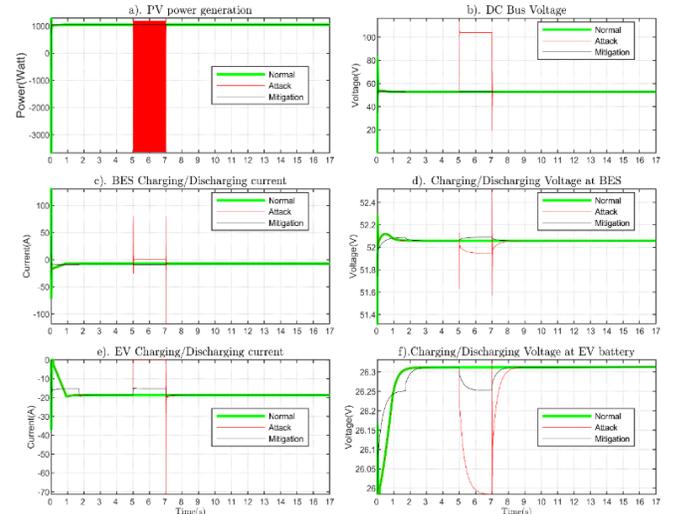

Fig.10. Impacts of Type II attack launched at PV controller from 5-7 seconds, BES controller from 5-7 seconds, and EV controller from 5-7 seconds and the mitigation performance during the attack.

Table X presents the features of the proposed mitigation methods with respect to other related works. Our proposed air-gapped TD3-based mitigation has surpassed the various state-of-the-art methods in attack detection with online mitigation with embedded intelligence.

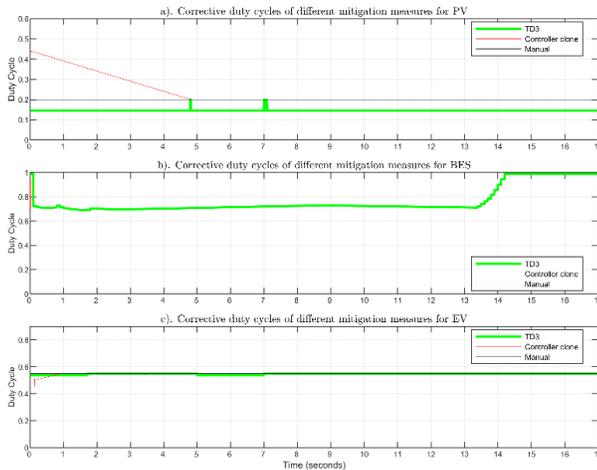

Fig.11. Control actions of Brute force, Controller clone, and TD3 mitigations

TABLE X
COMPARISON BETWEEN THE PROPOSED AND THE STATE-OF-THE-ART ALGORITHMS

| Solution | Attack detection | Coordinated Attacks | Online Mitigation | Embedded Intelligence | Air gapped |
|---|---|---|---|---|---|
| TD3 (our work) | √√ | √√ | √√ | √√ | √√ |
| Clone (our work) | √√ | √√ | √√ | X | √√ |
| Bruteforce (our work) | √√ | √√ | √√ | X | √√ |
| HIDS for EVCS [13] | √√ | √√ | X | √√ | √√ |
| NIDS for EVCS [7] | √√ | √√ | X | X | X |
| Weighted attack defense tree [10] | √√ | √√ | X | X | X |

## VII. CONCLUSION

This work devised three different mitigation techniques such as manual/brutefoce mitigation, controller clone-based mitigation, and DRL-based TD3 mitigation against the APT-based cyberattacks on EVCS controllers. The case studies uncover the following findings:

- The repetitive low-frequency attack (Type-I) on all controllers, at different times or simultaneously, has adverse impacts on critical functionalities of all controllers with the tendency to damage the EVCS with an upsurge/down surge in electrical signals. The agents successfully restore the EVCS operation by correcting the control signals of legacy controllers.
- The constant attack (Type-II) on controllers at different times or simultaneously tends to corrupt and damage the electrical components related to the legacy control actions. The proposed agents attempt to correct the control signals with the least error.
- The proposed Bruteforce-based mitigation is simple, but it can't adapt to dynamics change. The controller clone can restore the operation; however, it shares the same vulnerabilities as the legacy controllers. So adaptive and intelligent MADRL-based TD3 agents could be the appropriate defense option.

Future research will focus on exploring other novel methods for detection and mitigation of cyberattacks on the EVCS controllers, and their performance will be compared with those in this work. Also, 5G based-communication [13] application to EVCS and related cybersecurity issues will be analyzed.